\documentclass{moriond}

\def\Journal#1#2#3#4{{#1} {\bf #2}, #3 (#4)}

\def\be{\begin{equation}}
\def\ee{\end{equation}}
\def\bea{\begin{eqnarray}}
\def\eea{\end{eqnarray}}

\newcommand{\dau}{\mbox{$d$$+$Au}}
\newcommand{\pau}{\mbox{$p$$+$Au}}
\newcommand{\hau}{\mbox{$^3$He$+$Au}}

\newcommand{\Photo}{}

\begin{document}
\vspace*{4cm}
\title{COLLECTIVITY IN RHIC GEOMETRY SCAN AS SEEN BY PHENIX}

\author{ T. Nov\'ak for the PHENIX Collaboration }

\address{EKU KRC\\
Gy\"ongy\"os, M\'atrai \'ut 36, Hungary}

\maketitle\abstracts{
In this paper 
we show azimuthal particle correlations in three different
small-system collisions with different intrinsic initial geometries. The
simultaneous constraints of $v_2$ and $v_3$ in $p/d/^3$He$+$Au collisions
definitively demonstrate that the $v_n$'s are correlated to the initial
geometry.
In addition, we find that hydrodynamical models which include
QGP formation describe  simultaneouly the elliptic and triangular flow data in a statistically acceptable manner in all three systems. }

\section{Introduction}

One of the key discoveries at RHIC is the identification of quark-gluon plasma (QGP) and its characterization as a near-perfect fluid via its collective flow~\cite{Arsene:2004fa,Back:2004je,Adams:2005dq,Adcox:2004mh,Heinz:2013th}.
One of the first observations of collective longitudinal and radial flow and their hydrodynamical coupling in the invariant momentum distribution and
Bose-Einstein correlations was made by the EHS/NA22 experiment ~\cite{Agababyan:1997wd} in $h$+$p$ collisions at CERN SPS at the beam momentum of 250 GeV/c, corresponding to $\sqrt{s} \approx 22$ GeV.
As one of the first results of the d+Au beam energy scan at RHIC, PHENIX observed collective hydrodynamical behaviour of elliptic flow in d+Au collisions~\cite{Aidala:2017pup,Aidala:2017ajz}, providing evidence for collectivity in d+Au collisions  from $\sqrt{s_{NN}} = 20$ GeV to 200 GeV.
The LHC experiments observed  similar features in small-system
collisions~\cite{Khachatryan:2010gv,CMS:2012qk,Abelev:2012ola,Aad:2012gla}.
These results not only broaden the domain of the applicability of the hydrodynamical paradigm to a previously unexpected domain, 
but also raise several fundamental questions as well. Is it due to the appereance of sQGP (i.e.~a strongly coupled fluid)? If yes, how much time is spent in the QGP phase? What is the origin of final state collectivity? Is it due to initial geometry and hydrodynamics? Is the initial state geometry the primary driver of final state momentum correlations in small systems?

In order to test and answer these questions RHIC performed not only beam energy scan but also geometry scan measurements which allows for the investigation of the phase diagram of QCD matter by varying the beam energy in the region where the change from crossover to first order phase transition is suggested to occur. The beam-energy-scan program found real-valued $v_2$ in $d+$Au at all collision energies, providing evidence for collectivity in $d+$Au at all energies. Applying the unique capabilities of RHIC a projectile geometry scan~\cite{Nagle:2013lja} was utilized in order to discriminate between hydrodynamical models that couple to the initial geometry and initial-state momentum correlation models that do not. 

To characterize the fluidity of QGP,
the azimuthal distribution of each event's final-state particles, $\frac{dN}{d\phi}$, is 
decomposed into a Fourier series as follows:
$   \frac{dN}{d\phi} \propto 1+ \sum_n 2 v_n(p_T)\cos(n(\phi-\psi_n)),$
where $p_t$ and $\phi$ are the transverse momentum and the azimuthal angle of a 
particle relative to the beam direction, respectively, and $\psi_n$ is the orientation of the $n^{\rm th}$ 
order symmetry plane of the produced particles.  The second ($v_2$) and third ($v_3$) Fourier coefficients 
represent the amplitude of elliptic and triangular flow, respectively. 

\begin{figure*}[htbp]
  \begin{center}
    \includegraphics[width=0.8\linewidth]{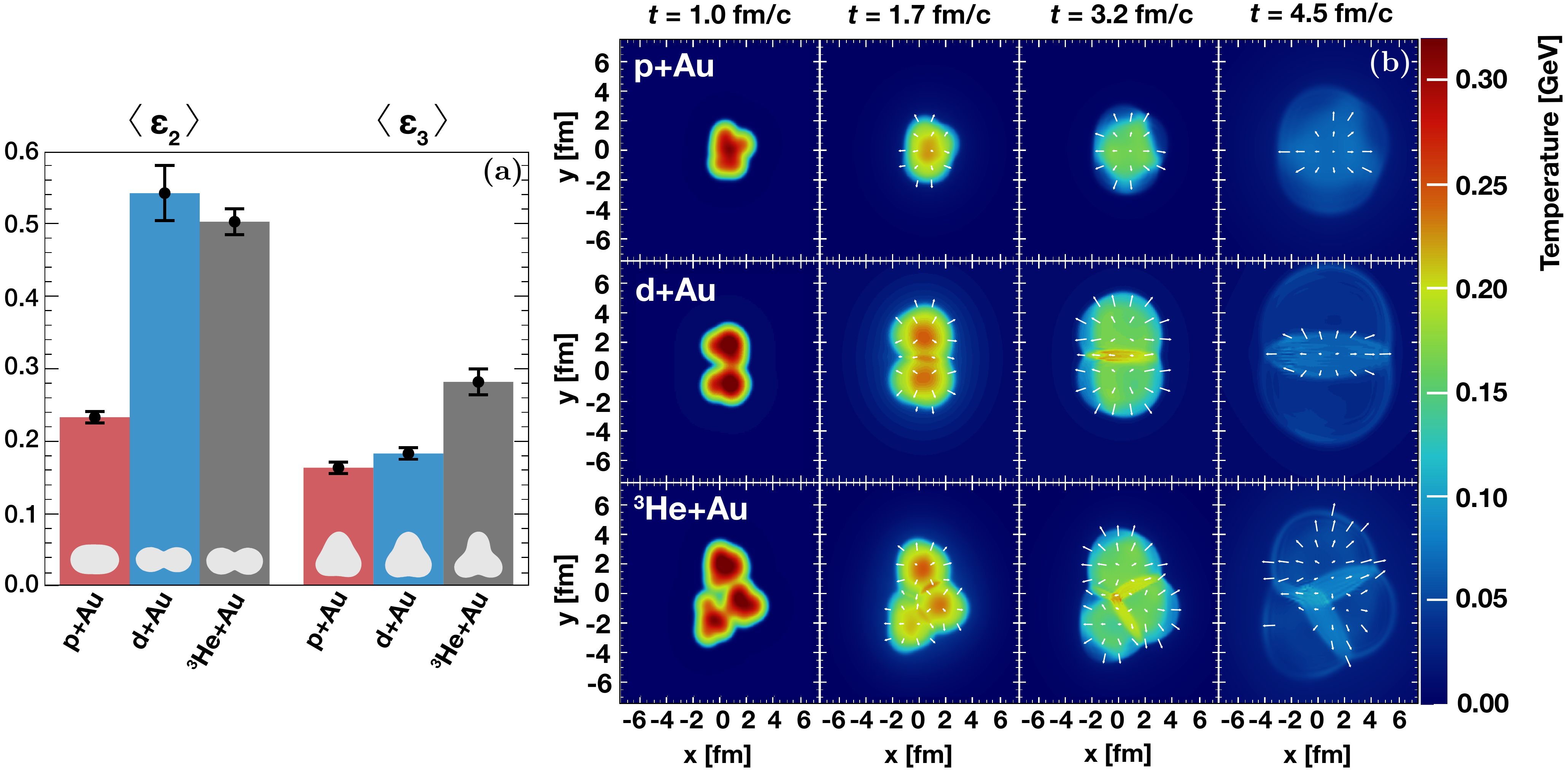}
    \caption{{\bf Average $\epsilon_n$ from a MC 
            Glauber model and hydrodynamic 
            evolution of small systems.}  }
    \label{fig:hydro}
  \end{center}
\end{figure*}

Varying the collision system from \pau, to \dau, to $^{3}$He+Au changes the initial 
geometry from dominantly circular, to elliptical, and to triangular 
configurations, as characterized by the 
2$^{\rm nd}$ and 3$^{\rm rd}$ order spatial eccentricities, which correspond to 
ellipticity and triangularity, respectively. The mean $\varepsilon _2$ and $\varepsilon _3$ values 
for small impact parameter $p/d/^{3}$He+Au collisions are shown in Fig.~\ref{fig:hydro}a. The definition of the $n^{\rm th}$ order spatial 
eccentricity of the system, $\varepsilon_n$, is
$    \varepsilon_n=\frac{\sqrt{\langle r^n\cos(n\phi)\rangle^2+\langle r^n \sin
    (n\phi)\rangle^2}}{\langle r^n \rangle},$
where $r$ and $\phi$ are the polar coordinates of participating 
nucleons~\cite{Alver:2010gr}. Based on the calculation from a MC Glauber model, the average second and third order spatial eccentricities ($\epsilon_2$ and $\epsilon_3$) are shown as columns in Fig.~\ref{fig:hydro}a. The second and third order spatial eccentricities are called  ellipticity and triangularity, respectively.

Hydrodynamical models begin with an initial spatial energy-density distribution with a given temperature that evolves in time following the laws of
relativistic viscous hydrodynamics using an equation of state determined from lattice QCD~\cite{Gale:2013da}. 
Examples of this temperature evolution are shown for
$p/d/^{3}$He+Au collisions in Fig.~\ref{fig:hydro}b using the hydrodynamical model SONIC~\cite{Habich:2014jna}. Based on haydrodynamical models a clear prediction for the ordering of the experimentally accessible $v_2$ and $v_3$ can be given, namely
  \begin{equation}
      v_2^{\pau}<v_2^{\dau}\approx v_2^{\hau}, 
  		\hspace{50pt}
      v_3^{\pau}\approx v_3^{\dau}<v_3^{\hau}.
      \label{eq:vn-hydro}
  \end{equation}
This ordering assumes that hydrodynamics can efficiently translate the initial 
geometric $\varepsilon_n$ into dynamical $v_n$, which is indeed seen in hydrodynamical simulations with small values of specific shear viscosity, as indicated on Fig.~\ref{fig:hydro}. 

There exist a class of alternative explanations where $v_n$ is not generated via flow, but rather is created at the earliest time in the collision process as described by so-called color glass condensate or initial momentum space correlation models~\cite{Mace:2018vwq}. The expectation from models based on initial-state momentum domain correlations  for the ordering of the magnitude of the $v_2$ and $v_3$ coefficients is:
  \begin{equation}
      v_n^{\pau}>v_n^{\dau}>v_n^{\hau},
      \label{eq:vn-momentum}
  \end{equation}
while  the MSTV model in which  gluons from the Au target do not resolve the individual color domains in the projectile  $p/d/^{3}$He does not follow Eq.~(\ref{eq:vn-momentum}).\footnote{Please see the Note Added in Proof at the end of this manuscript for an important update regarding the MSTV calculation.}

\section{Models vs.~data}

\begin{figure*}[htbp]
  \begin{center}
    \includegraphics[width=0.75\linewidth]{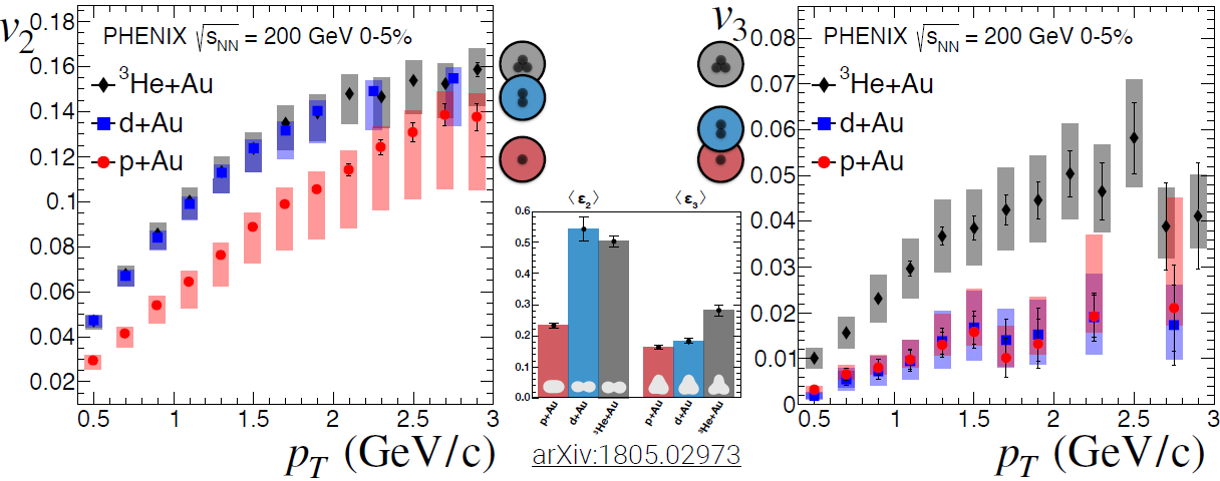}
    \caption{ {\bf PHENIX results for $v_2(p_t)$ and $v_3(p_t)$ in the RHIC geometry scan at $\sqrt{s_{NN}} = 200$ GeV.}  }
    \label{fig:v2_v3}
  \end{center}
\end{figure*}

Fig.~\ref{fig:v2_v3} summarizes the results of elliptic and triangular flow measurements in the RHIC 
$p/d/^{3}$He+Au 
geometry scan. The data points follow a geometrical ordering in a qualitative agreement with expectations from hydrodynamics.

\begin{figure*}[htbp]
  \begin{center}
    \includegraphics[width=0.75\linewidth]{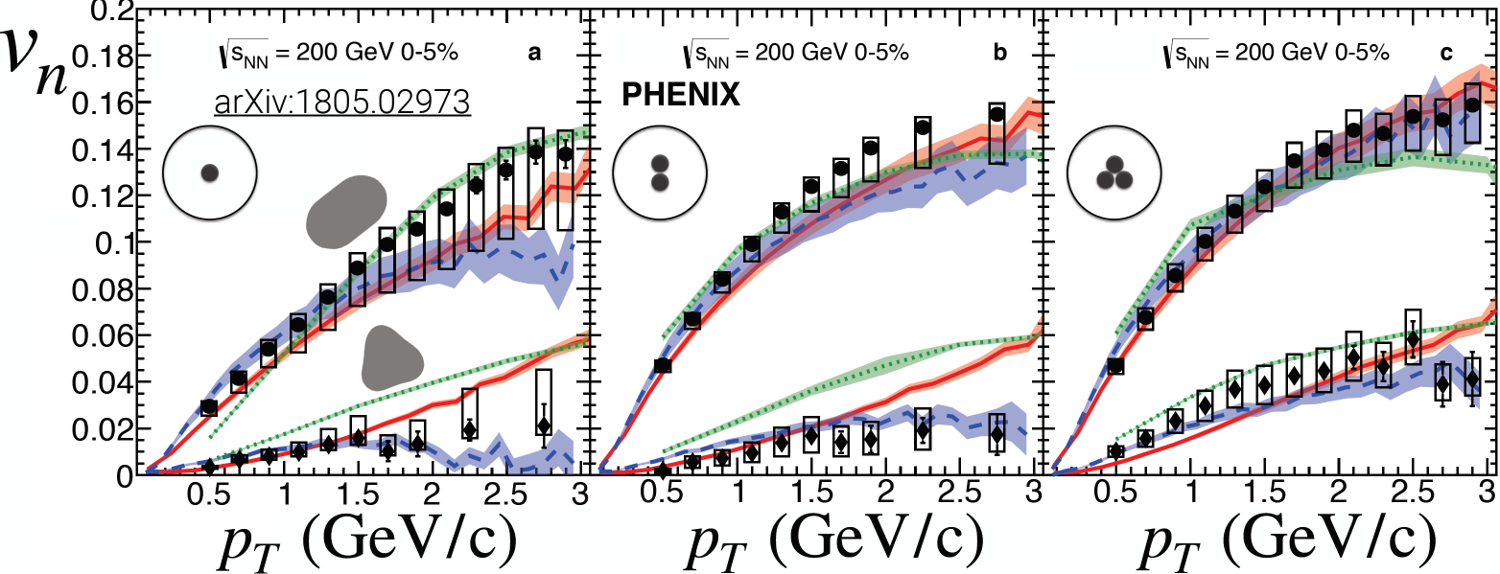}
    \caption{{\bf Elliptic and triangular flows as a function of $p_T$ in the RHIC geometry scan.} Panel a) shows results for  p+Au, panel b) for d+Au and panel c) for $^{3}He+Au$ collisions at $\sqrt{s_{NN}} = 200$ GeV in 0-5\% centrality class, as compared to SONIC (solid red) , VISHNU (dashed blue) predictions and MSTV (solid green) postdictions.} 
    \label{fig:hydro_mstv}
  \end{center}
\end{figure*}

Fig.~\ref{fig:hydro_mstv} compares quantitatively the PHENIX elliptic and triangular flow measurements for $p/d/^{3}$He+Au collisions with the results of numerical simulations. Two of these, SONIC and iEBE-VISHNU indicate predictions from numerical solutions of 2d+1 relativistic hydrodynamics with lattice QCD equation of state. The third model MSTV is on the other hand is based on initial state correlations and a color glass condensate initial state.
 Hydrodynamical models are consistent with the $v_n$ data in all three systems, however, they tend to diverge at higher $p_T$ in case of $v_3$, which may be more sensitive to the hadronic scattering. Focusing on the MSTV, Fig.~\ref{fig:hydro_mstv} shows that this model does a fair job in case of $v_2$, but fails in case of $v_3$. 

In order to distinguish these models, a statistical significance test was made and provided a $p$-value for the MSTV calculations of $v_2$ and $v_3$ for the three collision systems of effectively zero, in contradiction to the robust values found for the hydrodynamical models. 

\begin{figure*}[htbp]
  \begin{center}
    \includegraphics[width=0.4\linewidth]{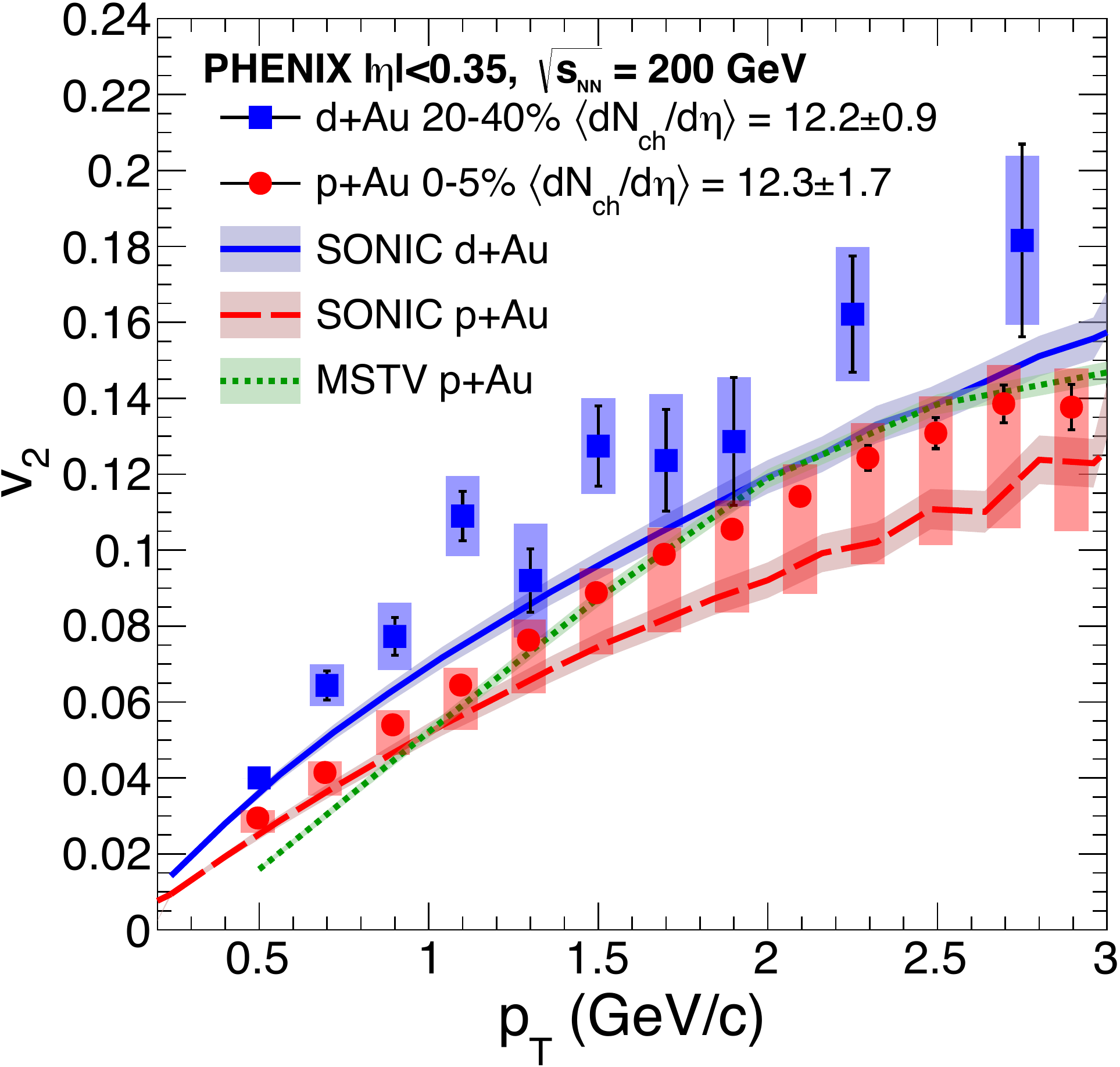}
    \caption{ {\bf Measured $v_2(p_T)$ in $p$+Au and $d$+Au collisions at the same event multiplicity, as compared to hydrodynamical calculations with SONIC and MVST color glass condensate calculations }(note that these calculatios predict the same green line for p+Au and d+Au collisions). }  
    \label{fig:samemult}
  \end{center}
\end{figure*}

The MSTV paper made a clear prediction that the $v_2$ will be identical between systems when selecting on the same event multiplicity. Shown in Fig.~\ref{fig:samemult} are the previously published \dau (20-40\%)
and \pau (0-5\%) $v_2$ where the measured mean charged particle multiplicities ($dN_{\mathrm{ch}}/d\eta$)
match~\cite{Adare:2018toe}. 
Our results contradict to this MSTV prediction, as they indicate clear differences between the $v_2$ of $d$+Au and $p$+Au collisions even if they are measured in the same multiplicity class, as indicated by Fig.~\ref{fig:samemult}. The results are however in a reasonable qualitative agreement with  hydrodynamical predictions.

\section*{Note Added in Proof}

Subsequent to the preparation of this manuscript we were made aware that there is an issue in the MSTV calculation and that the calculation no longer agrees with the PHENIX data when the issue is corrected. For details see 
\url{http://www.int.washington.edu/talks/WorkShops/int_19_1b/People/Mace_M/Mace.pdf} .

\section*{Acknowledgments}

The author is grateful for the support of EFOP 3.6.1-16-2016-0001, and NKFIH grant FK 123842 - 123959 (Hungary), as well as to the full list of \href{https://www.bnl.gov/rhic/phenix.asp}{PHENIX funding agencies}.


\section*{References}

\begin{thebibliography}{99}

\bibitem{Arsene:2004fa}
Arsene, I. \emph{et~al.} \Journal{Nucl. Phys. A}{757}{1}{2005}.

\bibitem{Back:2004je}
Back, B.~B. \emph{et~al.}
\Journal{Nucl. Phys. A}{757}{28}{2005}.

\bibitem{Adams:2005dq}
Adams, J. \emph{et~al.}
\Journal{Nucl. Phys. A}{757}{102}{2005}.

\bibitem{Adcox:2004mh}
Adcox, K. \emph{et~al.} \Journal{Nucl. Phys. A}{757}{184}{2005}.

\bibitem{Heinz:2013th}
Heinz, U. \& Snellings, R.
\Journal{Ann. Rev. Nucl. Part. Sci.}{63}{123}{2013}.

\bibitem{Agababyan:1997wd} 
  N.~M.~Agababyan {\it et al.} [EHS/NA22 Collaboration],
  Phys.\ Lett.\ B {\bf 422}, 359 (1998)

\bibitem{Aidala:2017pup} 
  C.~Aidala {\it et al.} [PHENIX Collaboration],
  Phys.\ Rev.\ C {\bf 96}, no. 6, 064905 (2017)

\bibitem{Aidala:2017ajz} 
  C.~Aidala {\it et al.} [PHENIX Collaboration],
  Phys.\ Rev.\ Lett.\  {\bf 120}, no. 6, 062302 (2018)


\bibitem{Khachatryan:2010gv}
Khachatryan, V. \emph{et~al.}
\Journal{ J.  High Energy Phys.} {\bf 09}, 091 (2010).

\bibitem{CMS:2012qk}
Chatrchyan, S. \emph{et~al.}
\Journal{Phys. Lett. B}{718}{795}{2013}.

\bibitem{Abelev:2012ola}
Abelev, B. \emph{et~al.}
\Journal{Phys. Lett. B}{719}{29}{2013}.

\bibitem{Aad:2012gla}
Aad, G. \emph{et~al.}
\Journal{Phys. Rev. Lett.}{110}{182302}{2013}.

\bibitem{Nagle:2013lja}
Nagle, J.~L. \emph{et~al.}
\Journal{Phys. Rev. Lett.}{113}{112301}{2014}.
  
\bibitem{Alver:2010gr}
Alver, B. \& Roland, G.
\Journal{Phys. Rev. C}{81}{054905}{2010}.

\bibitem{Gale:2013da}
Gale, C.,Jeon, S. \& Schenke, B.
\Journal{Int. J. Mod. Phys. A}{28}{1340011}{2013}.
  
\bibitem{Habich:2014jna}
Habich, M., Nagle, J.~L. \& Romatschke, P.
\Journal{Eur. Phys. J. C}{75}{15}{2015}.

\bibitem{Mace:2018vwq}
Mace, M., Skokov, V.~V., Tribedy, P. \& Venugopalan, R.
\Journal{Phys. Rev. Lett.}{121}{052301}{2018}.


\bibitem{Adare:2018toe}
Adare, A. \emph{et~al.}
{1807.11928}{~(2018)}.

  
 

\end{thebibliography}
\end{document}